\documentstyle[12pt]{article}
\newlength{\extralineskip}
\newcommand{\beq}{\begin{equation}}
\newcommand{\eeq}{\end{equation}}
\newcommand{\bd}{\begin{displaymath}}
\newcommand{\ed}{\end{displaymath}}
\def\bea{\begin{eqnarray}}
\def\eea{\end{eqnarray}}
\def\nn{\nonumber}
\def\ba{\beq\new\begin{array}{c}}
\def\ea{\end{array}\eeq}

\def\inbar{\,\vrule height1.5ex width.4pt depth0pt}
\def\IC{\relax\hbox{$\inbar\kern-.3em{\rm C}$}}
\def\IR{\relax{\rm I\kern-.18em R}}

\def\1{\relax{\rm 1\kern-.25em l}}

\def\Tr{{\rm Tr}}
\def\e{~{\rm e}}
\def\la{{\langle}}
\def\ra{{\rangle}}
\makeatletter
\newdimen\normalarrayskip              % skip between lines
\newdimen\minarrayskip                 % minimal skip between lines
\normalarrayskip\baselineskip
\minarrayskip\jot
\newif\ifold             \oldtrue            \def\new{\oldfalse}
\begin{document}
\begin{titlepage}
\thispagestyle{empty}
 
\begin{flushright}
PUPT-1893 
\end{flushright}

\vskip1cm
\begin{center}
{\LARGE \bf Chern-Simons Gravity, Wilson Lines and Large N Dual Gauge Theories}\\
\vskip1cm
{\bf L.D. Paniak} \\
\bigskip
{\it Joseph Henry Laboratories, Department Of Physics \\
Princeton University \\
Princeton, New Jersey 08544, USA} \\
e-mail: paniak@feynman.princeton.edu
\vskip4cm

\begin{abstract}

A five-dimensional Chern-Simons gravity theory based on the anti-de Sitter
group $SO(4,2)$ is argued to be 
a useful model in which to understand the details of 
holography and the relationship between generally covariant
and dual local quantum field theories.  Defined on a manifold with 
boundary, conformal geometry arises
naturally as a gauge invariance preserving 
boundary condition. By matching thermodynamic quantities for
a particular background geometry,
the dimensionless coupling constant of the Chern-Simons theory
is directly related to the number of fields in a putative 
dual theory at high temperature. 
As a consistency check, 
the semiclassical factorization of Wilson line observables in the 
gravity theory is shown to induce a factorization in dual theory 
observables as expected by general arguments of large $N$ gauge theory.
\end{abstract}

\end{center}
\end{titlepage}
\newpage
\setcounter{page}1
 
Holography in $D$-dimensional generally covariant system 
is the general proposition which states that the number of degrees 
of freedom in the system is bounded from above by the 
area (in Planck units) of a $D-2$-dimensional spatial surface surrounding the 
system \cite{thooft,suss,bousso}.  An ambitious extension of 
this concept is the organization of the so-called boundary 
degrees of freedom into a local quantum field theory
which effectively describes the physics of the $D$-dimensional
system.
The existence of such `dual' theories to certain generally
covariant systems has been supported by much recent 
work, especially  in the case of compactified string 
theory and its relation to supersymmetric gauge theory.
Initial evidence from D-brane scattering calculations
\cite{kleb1,kleb2} has been formalized in 
the `AdS/CFT' correspondence principle
\cite{malda1,gkp,witten3} (see \cite{rev} for a 
comprehensive review with exhaustive reference list).
Unfortunately,  due to the complexities of explicit
calculations on both the string and gauge theory sides
of the duality, the details of the relationship between
these theories remain uncertain.

Since holography is considered to be a generic feature of 
generally covariant systems and it can be argued 
theories which admit anti-de Sitter space-times also admit
dual interpretations \cite{susswitt}, it may be useful to examine examples
outside of string theory with these properties.
With these conditions in mind, we propose
to examine a generally covariant system in five space-time
dimensions which is based on a pure Chern-Simons gauge theory. 
With the particular 
choice of the anti-de Sitter group $SO(4,2)$ as the 
gauge group, the Chern-Simons theory defines a higher curvature generalization
of Einstein gravity \cite{chams1,chams2,bgh} which naturally admits
anti-de Sitter space-times. In addition, when defined on space-times
with a boundary, the natural gauge invariance preserving boundary 
condition of a flat connection immediately leads to a boundary 
manifold on which unrestricted gauge transformations generate conformal symmetries.
In this way, the Chern-Simons gravity theory is similar to low-energy
string theory effective actions which admit anti-de Sitter backgrounds with 
conformally invariant boundaries, but there are differences.
Unlike higher curvature terms  which
are expected due to stringy $\alpha^\prime$ corrections to Einstein gravity, 
in Chern-Simons gravity higher curvatures are not suppressed relative
to the linear curvature term.  Consequently,  
the Chern-Simons gravity theory we will discuss here 
does not follow in an obvious way as the low energy effective 
description of a string theory configuration. Unfortunately this
means we will not have at our disposal the well-known machinery to 
identify the conjectured dual gauge theory.  Moreover, since 
Chern-Simons gravity is not renormalizable in five space-time
dimensions we will restrict ourselves to the semiclassical 
regime and consider the theory in the vicinity of 
solutions of the classical equations of motion. 

While it may seem 
paradoxical to use a gauge theory to define a gravitational
theory, the Chern-Simons gravity theory we will consider is
a direct generalization of the $2+1$-dimensional theory
which describes Einstein gravity in three dimensions
\cite{at,witten1}.  As in the three-dimensional 
case, the relationship to a gauge theory leads to 
the most intriguing property of a Chern-Simons formulation of 
gravity - Wilson line observables.  These are natural 
gauge-invariant observables in the Chern-Simons theory and are carried 
over to the gravity theory as a concrete and effective
definition of non-local, diffeomorphism-invariant observables. 
Assuming there
exists a dual (holographic) theory which describes the same physics as
the gravitational theory, such observables should
have a natural interpretation in terms of observables of a local 
quantum field theory.
\beq
\la X \ra_{CS} = \la Y \ra_{QFT}
\eeq
Here we will show explicitly  this relationship implies the 
factorization of field theory observables consistent with 
the expectations for large rank gauge theories.
This will be accomplished in the semiclassical regime of 
the Chern-Simons gravity theory by finding a 
specific background geometry which is argued to control 
the system at high temperatures. The matching 
of free energies between the gravitational theory and a 
hot, rank $N$ gauge theory leads to an 
identification between the number of gauge degrees of freedom and the coupling 
constant of the Chern-Simons gravity theory.  In this way, the 
large $N$ limit of the gauge theory is shown to correspond to 
the semiclassical limit of the gravity theory and the factorization 
of observables follows.

Chern-Simons gravity is a higher curvature generalization 
of Einstein gravity \cite{chams1,chams2,bgh} based on a five-dimensional pure Chern-Simons theory
with anti-de Sitter gauge group, $SO(4,2)$.  The five-dimensional Chern-Simons form 
is defined via the relation
\beq 
d \omega_{CS}=   \la  F  \wedge F \wedge F  \ra
\eeq
Here $F$ is the curvature two-form related to the  gauge connection $A$
in the familiar way
\beq
F  = d A  + A  \wedge A 
\eeq
The bracket $\la \cdots \ra$ defines an invariant trace operation on the gauge algebra.
In general, it is not unique and, in particular,
depends on the representation of the gauge group chosen.
The choice which will be used in the sequel is that of the 
fully anti-symmetric tensor invariant of $SO(4,2)$
\beq
\la J_{B_1 C_1} J_{B_2 C_2} \cdots J_{B_n C_n} \ra
= \epsilon_{B_1 C_1 B_2 C_2 \cdots B_n C_n} 
\eeq
Here $J_{BC}$ are generators of the gauge group and 
$B,C= 1, \ldots,6$  label $so(4,2)$ algebra indices.
The Chern-Simons form can be explicitly calculated in terms of the
connection
\beq
\omega_{CS} = \la A \wedge dA \wedge dA + \frac{3}{2} A \wedge A \wedge 
A \wedge dA +\frac{3}{5} A \wedge A \wedge A \wedge A \wedge A \ra
\label{csform}
\eeq
The action is simply the integral of this five-form over
a given manifold ${\cal M}$
\beq
S= k \int_{\cal M} \omega_{CS}
\label{csaction}
\eeq
where $k$ is a dimensionless coupling constant.
We will restrict ourselves to topologically trivial 
${\cal M}$ for simplicity.

Of primary interest are manifolds with boundary.
In order  to maintain gauge invariance of the action on a manifold with 
boundary, it is necessary to impose boundary conditions on the 
gauge fields and gauge transformations.
Under a gauge transformation 
\beq
A_g = g^{-1} A g + g^{-1} d g
\eeq
the Chern-Simons form transforms as
\beq
\omega_{CS} (A_g) =  
\omega_{CS} (A) + d \alpha(A,g) + \frac{1}{10} \la ( g^{-1} dg)^5
\ra 
\label{cstform}
\eeq
The last term in (\ref{cstform}) has the form of a winding number
and upon integration over ${\cal M}$ takes values in the fifth cohomology group
$H^5({\cal M}, \pi_5( SO(4,2)) )$.  Since $\pi_5( SO(4,2))$ is pure torsion
the cohomology group is trivial and there is no quantization 
of the coupling constant $k$.
Consequently, for manifolds without boundary the action is gauge
invariant.  With a boundary, there generically   is a
non-vanishing contribution from the four-form $\alpha$  
\beq
\alpha(A,g) = \la -\frac{1}{2} dg g^{-1}  ( A   F + F  A) 
+ \frac{1}{2} dg g^{-1}  A^3 +
\frac{1}{4} (dg g^{-1}  A )^2
+\frac{1}{2} (dg g^{-1})^3  A
\ra
\eeq
Gauge invariance requires the integral of $\alpha$ over the four-
dimensional boundary $\partial {\cal M}$ to vanish.  There are a number
of ways to achieve this result.  One way is to follow as in 
the three-dimensional case \cite{jones} and restrict the gauge transformations 
allowed at the boundary
to be trivial and leave the gauge field unconstrained. 
The restriction we will use here is different and
has an appealing interpretation in terms of a conformal geometry 
induced on the boundary.  We will constrain the curvature $F$ to vanish on the boundary 
which allows one to choose the connection $A$ to be pure gauge.
Gauge transformations  at the boundary are left unrestricted.  It can be shown 
that under this restriction the Chern-Simons form transforms
invariantly up to winding terms which we have previously 
argued are irrelevant for the gauge group $SO(4,2)$.

Having specified the action and associated boundary conditions on
the fields, the equations of motion follow from a bulk variation of the action 
(\ref{csaction}) and are conveniently given by 
\beq
\epsilon_{B_1 C_1 B_2 C_2 B_3 C_3} F^{B_1 C_1} \wedge F^{B_2 C_2} =0 
\label{classeqmo}
\eeq
On a manifold with boundary, the variational principle for an action 
is typically ill-defined due to boundary contributions.
In the present case the boundary contribution can be calculated  as
\beq
\int_{\partial {\cal M}} \la A \wedge A \wedge A \wedge \delta A \ra
\eeq
Fortunately, the choice of flat boundary
conditions can be argued to eliminate these contributions \cite{gegen}.

The relation of the Chern-Simons action to a theory of gravitation 
follows from decomposing the connection $A$ on a basis of the 
anti-de Sitter algebra $\{P_a, J_{ab} \}$. 
\beq
A= \frac{e^a }{\lambda} P_a + \omega^{ab} J_{ab}
\label{adsdecomp}
\eeq
The coefficients of this 
expansion are identified with the f\"{u}nfbein and spin connection
\beq
A^{ab} = \omega^{ab}~~,~~ A^{a5} = 
\frac{e^a}{\lambda}~~ \mbox{a,b =0,1, \ldots,4}
\eeq
An extra length scale parameter $\lambda$, related to the cosmological
constant, has been introduced in the definition of the f\"{u}nfbein.
These definitions lead to the following identifications for the gauge
field curvature
in terms of geometric curvature and torsion
\bea 
F^{ab} &= & d \omega^{ab} +\omega^{ac} \wedge \omega_{c}^{~b} + 
\frac{ e^a \wedge e^b}{\lambda^2}  
\equiv R^{ab} + \frac{ e^a \wedge e^b}{\lambda^2}
\label{fdecomp} \\
F^{a5} &= & de^a + \omega^a_{~c} \wedge e^c  \equiv T^a
\nn
\eea

The flatness condition $F=0$ from  the gauge theory translates to a 
restriction to torsion-free configurations with a specific curvature
two-form
\beq
R^{ab} +  \frac{e^a \wedge e^b}{\lambda^2} =0 ~~~,~~~T^a =0
\eeq
This curvature two-form describes a space with constant, negative
curvature - an anti-de Sitter
space - and we see our choice of gauge invariance preserving
boundary conditions $F=0$ produces the natural setting 
for the duality between a bulk gravitational theory and
some local quantum field theory associated with the 
boundary.  As is well-known by now, taking the boundary of the manifold 
to be defined as the surface on which the anti-de Sitter
metric becomes singular induces a conformal structure 
on the boundary $\partial {\cal M}$.  Moreover, the unrestricted
$SO(4,2)$ gauge freedom at the boundary acts on this 
space as conformal motions \cite{witten3}.
 
The result of substituting the decomposition of the gauge field 
(\ref{fdecomp}) into the Chern-Simons form (\ref{csform})
is a higher curvature generalization of the
Einstein-Hilbert action which has been investigated previously
\cite{chams1,chams2,bgh}
\beq
S = k \epsilon_{abcde} 
\int_{\cal M} \left[ \frac{ e^a \wedge R^{bc} \wedge R^{de}}
{ \lambda }
+\frac{2 e^a \wedge e^b \wedge e^c \wedge R^{de}  }{3 \lambda^3}
+\frac{ e^a \wedge e^b \wedge e^c \wedge e^d \wedge e^e }{5 \lambda^5} 
\right] 
\eeq
and the equations of motion in gravitational variables are given by
\bea
\epsilon_{abcde} ( R^{ab} + \frac{ e^a \wedge e^b}{\lambda^2} )
\wedge ( R^{cd} + \frac{ e^c \wedge e^d}{\lambda^2} )   & = &  0
\label{classeqns} \\
\epsilon_{abcde} ( R^{ab} + \frac{ e^a \wedge e^b}{\lambda^2} )
\wedge T^c   & = & 0 \nn
\eea

Since Chern-Simons theory in five dimensions is not renormalizable,
we restrict ourselves to the semiclassical approximation where the 
coupling constant $k$ is large.  
Hence, we will be interested in the behaviour of the theory in
the vicinity of solutions of the classical equations of motion.
Immediately the symmetry of the theory suggests a background which is 
consistent with the boundary condition, namely anti-de Sitter
space.  Unlike the more familiar cases of low energy approximations
to string theory, here the anti-de Sitter background is not physically appealing.
To see that the Chern-Simons gravity theory is pathological in this background 
one needs to look no further than the classical equations
of motion (\ref{classeqmo}) and recall anti-de Sitter space corresponds
to vanishing curvature in the gauge theory.  Since the equations of motion are
quadratic in the curvature $F$, all fluctuations about the anti-de Sitter 
background vanish identically.  In terms of an effective quantum theory 
about this background it means there are no propagating degrees of 
freedom and the theory is perturbatively sterile.  Moreover, when calculating 
observables of Chern-Simons gravity we would find they are all 
trivial (up to topology of the space-time).  In short, Chern-Simons gravity in 
the anti-de Sitter background cannot describe any non-trivial field theory.

Fortunately, other solutions of Chern-Simons gravity are known to exist.
Of most interest to the current situation are those which are analogous to 
asymptotically anti-de Sitter Schwarzschild black holes in Einstein gravity \cite{hp}.
These have been discussed in detail for Chern-Simons gravity in \cite{btz} and for
similar theories in \cite{lovsolns}.
In particular, we are interested in a black hole, with mass $M$ related to a 
dimensionless parameter $a$, defined by the line-element
\beq
ds^2 = - ( 1 - a + (r/\lambda)^2) dt^2 + \frac{ dr^2}{ 1 - a  + (r/\lambda)^2} + r^2 d \Omega^2
\label{btzsoln}
\eeq
where $d \Omega^2$ is the volume element of the three-sphere.  It is easily checked
that this solution of (\ref{classeqns}) leads to a flat Chern-Simons connection on the boundary 
at infinity radius in accordance with our boundary conditions.
Since this solution represents a black hole geometry, there is a natural 
temperature associated with the system which can be calculated by continuing to 
Euclidean time with the result \cite{btz,lovtds}
\beq
\beta =  \frac{1}{T} =\frac{2 \pi \lambda}{ \sqrt{ a -1}}
\eeq
As well, the Euclidean action of Chern-Simons gravity can be calculated in this background
relative to zero temperature and vanishing mass $M(a)$ at $a=1$ 
\beq
I_{E} = \beta M(a) - \frac{ 16 \pi^2 \lambda}{3 \beta}
( 3 + \frac{4 \pi^2 \lambda ^2}{\beta^2})
\label{eaction}
\eeq

Important to note is that as the temperature is increased, the 
contribution of thermal gravitons \cite{gh} to the action  in the 
second term of (\ref{eaction}) becomes arbitrarily negative.
Therefore, it seems reasonable to conclude for sufficiently 
high temperature that the background (\ref{btzsoln}) will dominate the
semiclassical calculation of the partition function. 
This is completely analogous to the domination of the partition function
at high temperatures by the Schwarzschild anti-de Sitter
black hole as investigated in quantum gravity by Hawking and Page \cite{hp}
and Witten \cite{witten3,witten4} in the context of the AdS/CFT correspondence.
Consequently, at high temperatures the partition function and free energy
of the system can be easily calculated semiclassically with the result
\beq
\beta F_{CS} = -\log{Z}  = k I_E
\label{gravF}
\eeq
 
If there exists a dual, local quantum field theory defined on the boundary 
of the five-dimensional space-time which is described
by Chern-Simons gravity in the background (\ref{btzsoln}) then we should compare 
the free energy to that of a high temperature quantum theory.
In general we expect the boundary theory to contain gauge fields and 
in the limit of high temperature we expect the system is in 
a (quark-gluon) plasma phase where fundamental degrees of freedom are
not confined.  If the gauge group is of rank $N$ there are order
$N^2$ gluonic degrees of freedom and possibly as many matter degrees of 
freedom.
An estimate of the free energy of such a system on a four-sphere of length scale $\lambda$
can be given by dimensional arguments
\beq
F_{QFT}  = - \sigma N^2 \lambda^3 T^4
\label{cftF}
\eeq
where $\sigma$ is a proportionality constant of order one.
Equating the free energies in (\ref{gravF}) and (\ref{cftF}), 
we find in the limit of high temperature
the Chern-Simons coupling constant is directly related to the 
number of degrees of freedom in the boundary field theory
\beq
k = \frac{3 \sigma}{64 \pi^4} N^2
\label{kident}
\eeq
This identification of the coupling constant with the number
of degrees of freedom is reminiscent of the work of \cite{troncoso} and
\cite{horava} 
on eleven-dimensional Chern-Simons gravity and M-theory.  In that case the
degrees of freedom were taken to lie in the gravitational 
theory, not its dual and the relation to the gravitational 
coupling constant followed from Mach's Principle. Like that case,
our arguments only hold if the number of degrees of freedom $N$ 
is large: Mach's Principle requires a uniform background and we have 
done all our calculations in the semiclassical limit of large $k$ 
which, by (\ref{kident}), is equivalent to large $N$.

A novel consistency check of the relationship between Chern-Simons gravity and 
an unknown, dual, local quantum field theory which
describes the same physics can be formulated in terms of the observables
of each theory.  Using the relationship (\ref{kident}), we will derive
the large $N$ factorization of gauge-invariant observables in the dual
theory from semiclassical gravity calculations.

While it is not understood in general how to define observables in a 
generally covariant theory,  Chern-Simons gravity contains
a natural family of observables in Wilson lines inherited from 
its gauge theory formulation
\beq
W_R[\gamma]=\Tr_R P \e^{i \oint_\gamma A_\mu dx^\mu}
=\Tr_R P \e^{i \oint_\gamma (e_\mu^a P_a/ \lambda + \omega^{ab}_\mu
J_{ab}) dx^\mu}
\label{genwloop}
\eeq
For finite-dimensional representations $R$ of $SO(4,2)$,
this expression is simply the trace of the holonomy 
around the closed path $\gamma$. Of course for a non-compact gauge 
group like $SO(4,2)$ there are infinite dimensional 
irreducible representations as well.  In this case, the
naive picture breaks down and the trace of the path ordered matrix
integral must be realized as a dynamical system of infinite degrees 
of freedom \cite{witten2,carlip}.  Physically, these observables
describe the trajectories of external test particles in the 
gravitational background. As in a compact gauge theory,
the invariant properties of these test particles are given 
in terms of the irreducible representations of the gauge
group which for $SO(4,2)$ are labelled by mass and spin.

The expectation value of a collection of Wilson lines
is given by the standard formula
\beq
\la W_{R_1}[\gamma_1] \cdots W_{R_n}[\gamma_n] \ra_{CS}
= \frac{1}{Z} \int {\cal D} A \e^{i k
\int_{\cal M} \omega_{CS}[A]} 
\Tr_{R_1} P \e^{i \oint_{\gamma_1} A_\mu dx^\mu} 
\cdots \Tr_{R_n} P \e^{i \oint_{\gamma_n} A_\mu dx^\mu} 
\label{wcorrel}
\eeq
If one assumes there exists a dual local quantum field
theory to the Chern-Simons gravity, this expectation
value has an alternate interpretation in terms of some gauge
invariant quantity, $V$ defined in the dual theory
\beq
\la W_{R_1}[\gamma_1] \cdots W_{R_n}[\gamma_n] \ra_{CS}=
\la V_{R_1, \ldots, R_n}[\gamma_1, \ldots \gamma_n] \ra_{QFT}
\eeq

Unfortunately, there are at least two difficulties with 
this identification in the present situation.  First, the Wilson lines in 
the Chern-Simons gravity theory, being observables in a 
generally covariant theory cannot depend on the geometry
of paths $\{ \gamma_i \}$.  They should only be sensitive to 
the topological properties of the paths.  This is not what
one expects from a dual local quantum gauge theory in which 
similar path-dependent objects would be generic.  In those cases one expects
the Wilson line would depend on the length, or enclosed 
area of the paths. 
Second, the calculation of such observables
in the Chern-Simons gravity theory appears to be beyond
reach in the general case and the conjectured dual
local field theory is completely unknown.  
 
Fortunately, in the semiclassical limit of large $k$ these
concerns are absent.  Leading contributions to expectation
values come from the vicinity of definite geometries
satisfying both the gauge invariance preserving 
boundary condition, $F=0$ and the classical equations of motion (\ref{classeqns}).
In this way the topological nature of the Wilson lines is broken
and we expect these observables (\ref{wcorrel}) to depend on geometric information.
In particular,
in the limit of high temperature we have argued the black hole
geometry (\ref{btzsoln}) dominates the semiclassical partition function.
This remains true when calculating the semiclassical contributions to 
the correlation function (\ref{wcorrel}) as well since, to leading order in large $k$,
the Wilson lines do not contribute to the saddle-point.
Evaluating (\ref{wcorrel}) in the background of the classical solution (\ref{btzsoln}),
the expectation 
value of Wilson lines in Chern-Simons gravity factorizes to leading order in 
large $k$
\beq
\la W_{R_1}[\gamma_1] \cdots W_{R_n}[\gamma_n] \ra_{CS} =
\la W_{R_1}[\gamma_1] \ra_{CS} \cdots  \la W_{R_n}[\gamma_n] \ra_{CS}
+ O \left( \frac{1}{k} \right)
\label{csloopfact}
\eeq
Consequently, the factorization of the gravitational
observables induces a factorization in the dual theory.
Identifying the Chern-Simons coupling constant $k$ with 
$N^2$ as in (\ref{kident}),  (\ref{csloopfact}) implies
\beq 
\la V_{R_1, \ldots, R_n}[ \gamma_1, \ldots \gamma_n] \ra_{QFT} = 
\la V_{R_1}[\gamma_1] \ra_{QFT} \cdots \la V_{R_n}[\gamma_n] \ra_{QFT}
+O \left( \frac{1}{N^2} \right)
\label{qftloopfact}
\eeq
which is the  factorization property of gauge invariant observables
in a large $N$ rank gauge theory, as expected 
from very general considerations.

The purpose of these brief arguments is to motivate further
investigation of the Chern-Simons gravity theory, and its supersymmetric
extension \cite{chams2,bgh}, for the
purposes of gaining a better understanding of the relationship
between generally covariant and quantum field theories.
First and foremost, one would like to know more details about the 
conjectured dual quantum theory. This information may be obtained
by a close investigation of explicit Wilson line calculations 
for the black hole background (\ref{btzsoln}) we have considered, in analogy
with calculations for ${\cal N}=4$ super Yang-Mills theory \cite{witten4}.
It may even be possible to deduce a `loop equation` for 
path-dependent objects in the dual theory from the 
Chern-Simons action.

\section*{Acknowledgements}
We would like to thank M. Costa, M. Gutperle, M. Henningson, 
I. Klebanov, L. Smolin and D. Waldram for discussions.
This work was completed with the   
support of the Natural Sciences and Engineering Research Council of Canada, 
NSF grant PHY98-02484 and the James S. McDonnell Foundation grant 
No. 91-48.

%\newpage 

\end{document}